\documentclass[10pt]{article}

\usepackage{a4wide}

\begin{document}

\def\O{{\cal O}}
\def\N{{\cal N}}
\def\>t{>_{\scriptscriptstyle{\rm T}}}
\def\enu{\epsilon_\nu}
\def\pint{\int{\d^3p\over(2\pi)^3}}
\def\gint{\int[\D g]\P[g]}
\def\nbar{\overline n}
\def\d{{\rm d}}
\def\e{{\bf e}}
\def\x{{\bf x}}
\def\y{{\bf y}}
\def\hn{{{\bf h}_n}}
\def\hm{{{\bf h}'_m}}
\def\hp{{{\bf h}_p}}
\def\0x{\x^\smalze}
\def\sperpx{{x_\perp}}
\def\sperpk{{k_\perp}}
\def\sbperpk{{{\bf k}_\perp}}
\def\sbperpx{{{\bf x}_\perp}}
\def\perpx{{x_{\rm S}}}
\def\perpk{{k_{\rm S}}}
\def\bperpk{{{\bf k}_{\rm S}}}
\def\bperpx{{{\bf x}_{\rm S}}}
\def\p{{\bf p}}
\def\q{{\bf q}}
\def\zr{{\bf z}}
\def\R{{\bf R}}
\def\A{{\bf A}}
\def\v{{\bf v}}
\def\u{{\bf u}}
\def\w{{\bf w}}
\def\U{{\bf U}}
\def\cm{{\rm cm}}
\def\l{{\bf l}}
\def\sec{{\rm sec}}
\def\Ckol{C_{Kol}}
\def\flux{\bar\epsilon}
\def\zq{{\zeta_q}}
\def\b{b_{kpq}}
\def\bun{b^{\scriptscriptstyle (1)}_{kpq}}
\def\bdu{b^{\scriptscriptstyle (2)}_{kpq}}
\def\z0q{{\zeta^{\scriptscriptstyle{0}}_q}}
\def\smalS{{\scriptscriptstyle {\rm S}}}
\def\smalze{{\scriptscriptstyle (0)}}
\def\smalI{{\scriptscriptstyle {\rm I}}}
\def\smalun{{\scriptscriptstyle (1)}}
\def\smaldu{{\scriptscriptstyle (2)}}
\def\smaltr{{\scriptscriptstyle (3)}}
\def\smalL{{\scriptscriptstyle{\rm L}}}
\def\smalD{{\scriptscriptstyle{\rm D}}}
\def\smal1n{{\scriptscriptstyle (1,n)}}
\def\smaln{{\scriptscriptstyle (n)}}
\def\smalA{{\scriptscriptstyle {\rm A}}}
\def\shell{{\tt S}}
\def\ball{{\tt B}}
\def\nav{\bar N}
\def\micron{\mu{\rm m}}
\font\brm=cmr10 at 24truept
\font\bfm=cmbx10 at 15truept
\centerline{\brm Coherence, space non-locality and lacunarity}
\centerline{\brm in a cascade model of turbulence}
\vskip 20pt
\centerline{Piero Olla$^1$ and Paolo Paradisi$^2$}
\vskip 5pt
\centerline{$^1$ISIAtA-CNR, 73100 Lecce Italy}
\centerline{$^2$FISBAT-CNR, 40129 Bologna Italy}
\vskip 20pt

\centerline{\bf Abstract}
\vskip 5pt
We use multiscale-multispace correlations and
Fourier transform techniques, to study some intermittent random 
field properties, which escape analysis by structure 
function scaling. These properties are parametrized in terms of a set of 
scale ratios, giving the typical interaction distances in space and 
scale of the random field fluctuations, and the characteristic lengths over 
which these fluctuations act coherently to generate intermittency. The
relevance of these techniques in turbulence theory is discussed.
\vskip 2cm
\centerline{\it Submitted to}
\centerline{\it Phys. Rev. Lett.}
\centerline{\it 4/17/98}
\vfill\eject
Turbulence is usually depicted as a situation of irregular fluid motion, with
energy being transferred from large to small scales, by interaction between eddies of increasingly smaller size. This mechanism of energy transfer, usually referred to with the name of cascade, takes different forms depending
on the choice of turbulence description. In closure \cite{lesieur}, a Fourier representation of the Navier-Stokes equation is typically used. In this case, the transfer of energy between eddies  is expressed in 
terms of an energy flux in wavevector space. This flux is a 
nonlinear kernel of the spectral energy density, and therefore a fully global quantity, carrying no information on spatial structure and transfer fluctuations. In intermittency modelling, instead, the description is necessarily local in space, at least implicitly. In the random beta model \cite{benzi85}, for example, this is obtained using a superposition of nested eddies of increasingly smaller size, with a fluctuating eddy to eddy energy transfer. 

In both circumstances, the dynamics forces correlations between the fluctuations at different scale and space locations. Only recently, however, there has been some interest in studying correlations in turbulence \cite{arneodo98}, \cite{benzi98}, beyond the standard approach of focusing on structure function scaling. 

A tool that is receiving great attention, especially in the study of intermittency, 
is synthetic turbulence \cite{eggers91}, \cite{benzi93}, \cite{juneja94}. Synthetic turbulence has no pretense of modelling turbulence dynamics, its goal being more the one of reproducing the statistics observed in experimental turbulence data. 
Even so, the multiplicative noise algorithms used to produce the turbulent signal contain implicit assumptions on the nature of the turbulent flow that go beyond the task of reproducing anomalous structure function scaling. 

Typically, these synthetic signals are expressed as a superposition of wavelets with random coefficients:
$$
\Psi(x)=\sum_n^N\sum_\hn A_\hn w_S(k_\hn,y_\hn,x),
\eqno(1)
$$
where $\Psi(x)$ is the synthetic turbulent signal, with $ w_S(k,y,x)$, wavelets
of scale $k$ and position $y$ \cite{daubechies}, and the vector index $\hn=(h_0,h_1,....h_n)$ identifying 
the position of the wavelet in the cascade through the sequence of its ancestors: the integer 
$h_n$ labels the $h_n$-th daughter wavelet generated at the $n$-th step in the cascade, by 
wavelet ${\bf h}_{n-1}$. The standard practice \cite{benzi93} is to arrange the wavelets on a tree 
structure in $k-y$ space, with $k_\hn\equiv\bar k_n=\bar k_02^n$, $h_i=\pm 1$ 
$y_\hn=y_{{\bf h}_{n-1}}+{h_n\over 2k_n}$. The amplitudes $A_\hn$ are then  generated through a random multiplicative process.

This picture closely parallels the one of a random beta model \cite{benzi85}; the resulting
correlation pattern has clearly a multifractal (multiaffine) nature,
depending on the singularity (lack of singularity) of $\Psi(x)$, as 
$N\to\infty$ in Eqn. (1). The difference, in the case of the 
beta model, is that the multiplicative random process,
arises explicitly out of the assumption of an energy transfer,   
local in both space and scale. 

The purpose of this letter is to present some techniques, based on the use
of wavelet and Fourier analysis, to extract informations on the algorithm 
of signal generation, directly from the signal statistics. 
We focus in particular on two important aspects: the determination of which 
''building block'' wavelet $w_S$ is being used in the synthetic signal, and 
which is the typical interaction distance between wavelets, as
expressed by the separation in $k-y$ space: $(k_\hn-k_{{\bf h}_{n-1}},y_\hn-y_{{\bf h}_{n-1}})$.
In real turbulence, these aspects have a direct interpretation in terms
of eddy structure and energy transfer properties.

The choice of the wavelet, identifies a cell of $k-y$ space, which acts coherently to generate
intermittency; this is in practice determined by the ''number of wiggles'' of the wavelet, i.e.
the product $a_S=k\lambda_S$ of its characteristic wavevector and spatial extension. We can thus focus on the case in which $w_S$ is a Gaussian wavepacket, without 
fear of loosing to much in generality:
$$
w_S(k,y,x)=\exp(ik(x-y)-(x-y)^2/\lambda^2_S).
\eqno(2)
$$
The interaction distance provides informations on the degree of non-locality of the dynamics, 
by telling us whether only nearest neighbor cells in $k-y$ space can interact directly, or 
longer range interactions are allowed. The algorithms of turbulence synthesis studied in 
\cite{benzi93}, \cite{arneodo98} and \cite{benzi98} are strictly local. To allow for non-locality,
the simplest approach is to treat the wavelet coordinates in $k-y$ space, on the same ground
as the amplitudes $A_\hn$, as stochastic variables. In this way, the wavelets take more the
aspect of eddies, with complete freedom of location (and overlap) in $k-y$ space \cite{olla98}.
The dynamics is described through the probability for the transition 
$(\ln A_{{\bf h}_{n-1}},\ln k_{{\bf h}_{n-1}},y_{{\bf h}_{n-1}})\to (\ln A_\hn,\ln k_\hn,y_\hn)$,
which is taken in the form: $p_A(A'/A|k'/k)p_k(k'/k)p_x(k|y-y'|)$, with $\int\d\ln y\,y^p$
$\times p_A(y|x)=c_px^{-\zeta_p}$ to insure power law scaling in $k$ for the moments of $A$.
The phases are supposed random so that intermittency in $\Psi$ arises simply as a consequence
of the intermittency of the amplitudes $A$.
We take for simplicity, $p_x$ and $p_k$ to be Gaussians respectively in space separation and
$\ln k_\hn/k_{{\bf h}_{n-1}}$. The transition probabilities over $n$ cascade steps will be
Gaussians as well:
$$
P_x(n,k|y-y'|)\simeq{k\over\pi^{1\over 2}\hat ba_S}\exp\Big(-{k^2|y-y'|^2\over\hat b^2a_S^2}\Big)
$$
$$
P_k(n;k'/k)={1\over(\pi n)^{1\over 2}\Delta z}
\exp\Big(-{(\ln(k'/k)-n\bar z)^2\over n\Delta z^2}\Big).
\eqno(3)
$$ 
with $\hat b$ giving the amount of space non-locality and 
$\Delta z/\bar z$ the degree of discreteness in scale, in the cascade algorithm. 
Thus, the wavelet wavevectors in the cascade will be centered  at $\bar k_n=\bar k_0\exp(n\bar z)$
with increasing width $\Delta\bar k_n\sim n^{1\over 2}\Delta z\bar k_0$. 
The probability $p_A$, which is the multiplier
distribution for the coeffincients $A_\hn$, can be obtained starting from the distribution
of the scaling exponents $\zeta_q$, using standard techniques \cite{chhabra89}. We do not
need its explicit form here, however.
\vskip 5pt
To measure effects such as the degrees of interaction non-locality and discreteness, in an 
intermittent signal of unknown origin, multiscale-multispace wavelet correlations 
would be clearly the instrument of choice \cite{arneodo98}, \cite{benzi98}. This is conditioned, however, to having
some knowledge of the structure of the building block wavelets $w_S$, especially, as
regards the
value of the parameter $a_S$.

The simplest wavelet correlations are $\langle|\Psi_{ky}|^2\rangle$ and
$\langle|\Psi_{ky}|^2|\Psi_{k'y'}|^2\rangle$,
where $\Psi_{ky}$ is the component of $\Psi$ on a wavelet $(-k)^{-s}\partial_x^sw_A(k,y,x)$.
Here, $w_A$ is taken again to be a Gaussian wavepacket, with $a_A=k\lambda_A$ not necessarily
equal to $a_S$, and $s$ is chosen large enough to kill large scale contributions to $\Psi$:
$$
\Psi_{ky}=\lambda_A^{-1}k^{-s}\int\d x\, w_A^*(k,y,x)\partial_x^s\Psi(x).
\eqno(4)
$$
In a cascade model like the one considered here, the wavelet components $\Psi_{ky}$ 
entering $\langle|\Psi_{ky}|^2$ $\times |\Psi_{k'y'}|^2\rangle$ may 
come from a single building block wavelet $w_S$ (''one-eddy'' contribution), if $|k-k'|$ 
and $|y-y'|$ are both small enough; otherwise they will come predominantly from two 
distinct $w_S$ placed at $(k,y)$ and $(k',y')$ (''two-eddy'' contribution). 
If $|y-y'|<L$, with $L$ the largest scales in the signal,
there will be a correlation between the two building block wavelets $w_S$, due to their
coming from some common ancestor at scale $\hat k\sim\hat ba_S|y-y|^{-1}$. This situation 
is expressed through the formula:
$$
\langle |A_\hn|^2|A_\hm|^2\rangle=c_4(k_\hm L)^{-\zeta_4}(k_\hn/k_\hm)^{-\zeta_2}
(k_\hm/k_\hp)^{\zeta_4-2\zeta_2}
\eqno(5)
$$
with $k_\hn>k_\hm$ and $p$ the cascade step at which the genealogical tree of $\hn$ and $\hm$ 
branches: $h_i=h'_i$ for $0\le i\le p$ and $h_i\ne h'_i$ for $i>p$. Thus, the lower the
branching takes place in the tree, the closer the correlation gets to its disconnected
limit $\langle |A_\hn|^2|A_\hm|^2\rangle\sim (k_\hm k_\hn)^{-\zeta_2}$.

Indicate with $C(k,y;k',y')$ the square modulus of the component of $w_S$ with respect to
$w_A$:
$$
C(k_A,y_A;k_S,y_S)=\left|\lambda_A^{-1}k_A^{-s}
\int\d x\,w_A^*(k_A,y_A,x)\partial_x^sw_S^*(k_S,y_S,x)\right|^2
$$
$$
\simeq {a_S^2\over a_A^2+a_S^2}
\exp\Big(-{2 k_A^2\over a_A^2+a_S^2}\Big(\Delta y^2+{a_A^2a_S^2\Delta k^2 \over 4k_
A^2}\Big)\Big),
\eqno(6)
$$
where $\Delta k=k_A-k_S$ and $\Delta y=y_A-y_S$. We can now write down the
expression for the second order correlation. From Eqn. (3), for $\ln kL >\bar z^3/\Delta z^2$, 
discreteness effects can be neglected, and we have, indicating with 
$\langle P_k(n,\bar k/\bar k_0)\rangle$, average over the size of the large scale
building block wavelets:
$$
\langle |\Psi_{ky}|^2\rangle=
c_2\int\d\bar y\d\ln \bar k\sum_n\bar k_n\langle P_k(n,\bar k/\bar k_0)\rangle (\bar
 kL)^{-\zeta_2}
C(k,y;\bar k,\bar y)
\simeq{c_2\pi a_S\over \bar za_A}(kL)^{-\zeta'_2}.
\eqno(7)
$$
where the difference $\epsilon=\zeta'_2-\zeta_2=\O(\bar z^{-3}\Delta z^2)$ is due 
to the fluctuation in the scale ratio between a daughter wavelet $\hn$ and her parent
${\bf h}_{n-1}$, and
we have used the result, valid for $\ln k/\bar k_0>\bar z^3/\Delta z^2$: 
$\sum_{n=1}^\infty (\bar k_n/k)P_k(n,k/\bar k_0)\simeq\bar z^{-1}(k/\bar k_0)^\epsilon$.
We have a similar expression for the one-eddy contribution to the 4-th order correlation:
$$
\langle |\Psi_{ky}|^2|\Psi_{k'y'}|^2\rangle_1
\simeq{c_4\pi a_S\over\bar za_A}(kL)^{-\zeta'_4}(k'/k)^{-\zeta'_2}C(k,y;k',y')
\eqno(8)
$$
From Eqns. (8) and (6), if the space-scale separation is large enough,
$\langle|\Psi_{ky}|^2|\Psi_{k'y'}|^2\rangle$ will be dominated by the two-eddy contribution. 
In this case,
however, it is not sufficient to impose that $kL$ be large, to avoid discreteness effects.
Limiting the calculation to the connected part of the correlation, we find:
$$
\langle |\Psi_{ky}|^2|\Psi_{k'y'}|^2\rangle_c
=2c_4\int_{\bar k'>\bar k}\d\ln\bar k\d\ln\bar k'\d\bar y\d\bar y'
\int_{\ln L^{-1}}^{\ln k}\d\ln\hat k\,
(\bar kL)^{-\zeta_4}(\bar k'/\bar k)^{-\zeta_2}
(\bar k/\hat k)^{\zeta_4-2\zeta_2}
$$
$$
\times\sum_n\sum_{m=0}^n\sum_{p=1}^m{\bar k_n\bar k_m\over\bar k_p}\langle P_k(
p,\hat k/\bar k_0)
\rangle P_k(n-p,\bar k/\hat k)P_k(m-p,\bar k'/\hat k)
$$
$$
\times P_x(\hat k|\bar y-\bar y'|)C(k,y;\bar k,\bar y)C(k',y';\bar k',\bar y').
\eqno(9)
$$
where $(\bar k_m/\bar k_p)P_x(\hat k|\bar y-\bar y'|)$ is the space density at $\bar y'$, of
wavelets ${\bf h}'_m$ generated from the branching at ${\bf h}_p$, given the presence
of a wavelet $\hn$ at $\bar y$, and $\bar k_n$ is the space density of wavelets $\hn$.
From here, informations on the space-scale structure of the intermittent signal can
in principle be obtained.  The slow decay of correlations between wavelets, shown in Eqn. (5), 
however, suggests that the effect of coherency, important for $k|y-y'|<a_S$, and the one of
interactions, important for $k|y-y'|<\hat ba_S$, will be superimposed in a way that is 
difficult to disentangle. This warns against trying to measure $a_S$ looking for crossovers 
in the space scaling of correlations. We have in fact two mechanisms of correlation 
distruction: one is the decay of the one-eddy contribution, due to the two analyzing wavelets 
ceasing to overlap with a single $w_S$. The second is the decay of the two-eddy contribution,
i.e. the decrease with distance of the correlation between different $w_S$.
Only the first effect, of course, gives informations about the shape of $w_S$.

Surprisingly, this strongly local information is more easily obtained using a
fully global instrument like a Fourier transform, at least if the random phase hypothesis
is satisfied. The reason is the maximal definition in scale of a Fourier component. 
In this case, a correlation like
$\langle \Psi_{-k}^2\Psi_{k-\Delta}\Psi_{k+\Delta}\rangle$, with $\Psi_k$ standard Fourier
components, will be non-zero only when $\Delta <\lambda_S^{-1}=k/a_S$; in fact, a 4-th
order correlation can receive contributions at most by two $w_S$ (plus their complex
conjugate), which cannot cover a section of $k$ including at the same time $-k$, $k-\Delta$
and $k+\Delta$, when $\Delta$ is large. In the example of our cascade model, we can calculate 
this fourth order correlation rather easily. In the case of negligible discreteness effects, 
we find:
$$
\langle \Psi_{-k}^2\Psi_{k-\Delta}\Psi_{k+\Delta}\rangle
\simeq{\pi^{5\over 2}a_S^4\delta(0)k^{-3-\zeta'_4}\over\bar z\alpha}
\Big[1+{4\pi k\over\e\hat ba_S\bar z^2\alpha\Delta}\exp\Big(-{a_S^2\Delta^2\over 4k^2}\Big)\Big]
\exp\Big(-{a_S^2\Delta^2\over 4k^2}\Big)
\eqno(10)
$$
where $\alpha=(2+\zeta'_2+a_S^2/2)^{1\over 2}$ and the two terms in square brackets are
respectively the one- and two-eddy contribution to the correlation.
\vskip 5pt
Once $a_S$ is known, it is possible to set $a_A=a_S$ in the analyzing wavelet, in order to 
maximize the overlap between the analyzing wavelets $w_A$ and the coherent regions 
identified by the building block wavelets $w_S$. At this point
it is possible to look at the dependence on space and scale separation, of the expression
for the correlation $\langle |\Psi_{ky}|^2|\Psi_{k'y'}|^2\rangle_c$ provided by Eqn. (9).
We consider first the case in which discreteness effects can be neglected, corresponding
to the regime $\ln k'/k>\bar z^3/\Delta z^2$. Setting $a_A=a_S$ we obtain:
$$
\langle |\Psi_{ky}|^2|\Psi_{k'y'}|^2\rangle_c\simeq
{2c_4\pi^{3\over 2}\over\bar z^3\hat ba_S}(kL)^{-\zeta'_4}(k'/k)^{-\zeta'_2}
\int_0^{\ln kL}\d x
\Big(1+A\e^{-2x}\Big)^{-{1\over 2}}
$$
$$
\times\exp\Big[ (\zeta'_4-2\zeta'_2)x-
\Big(\e^{2x}+F\Big)^{-1}{k^2|y-y'|^2\over\hat b^2a_S^2}\Big]
\eqno(11)
$$
where $F=\hat b^{-2}(1+(k/k')^2)$.
From inspection of this equation, we see that the integral receives contribution, for
$\hat ba_S$ large enough, from $\max(0,\ln{k|y-y'|\over\hat ba_S}),\ln F)<x<\ln kL$, where 
the integrand is essentially $\exp((\zeta'_4-2\zeta'_2)x)$. We find then, for 
$2\zeta'_2-\zeta'_4$ small:
$$
\langle |\Psi_{ky}|^2|\Psi_{k'y'}|^2\rangle_c\simeq
{2c_4\pi^{3\over 2}\over\bar z^3\hat ba_S}(kL)^{-\zeta'_4}(k'/k)^{-\zeta'_2}
\max\Big(\ln GkL,(2\zeta'_2-\zeta'_4)^{-1}\Big)G^{2\zeta'_2-\zeta'_4}
\eqno(12)
$$
where $G=\max(1,{k|y-y'|\over \hat ba_S},F^{1\over 2})$. We thus have a close range
and a large separation range, with transition at 
$|y-y'|\sim(a_S/k)\max(\hat b,(1+(k/k')^2)^{1\over 2})$. In the close range, the 
correlation scales only in $k$: $\langle |\Psi_{ky}|^2|\Psi_{k'y'}|^2\rangle_c\sim
(kL)^{-\zeta'_4}(k'/k)^{-\zeta'_2}\ln kL$, while, in the 
large separation regime, the correlation contains a factor scaling like a power 
in $|y-y'|$: $\langle |\Psi_{ky}|^2|\Psi_{k'y'}|^2\rangle_c\propto 
|y-y'|^{\zeta'_4-2\zeta'_2}$. This slow decay in $|y-y'|$, observed also in 
\cite{arneodo98}, and the one of structure functions, with respect to $a_A$ 
\cite{olla98}, have the same origin in the dependence on $k_\hp$ of 
$\langle |A_\hn|^2|A_\hm|^2\rangle$ $[$see Eqn. (5)$]$: larger space separations
between building block wavelets imply a lower branching in their genealogical tree.

Thus, if $\hat b$ is large, for any scale $k$ there is a second characteristic
length, beyond $\lambda_S$, which gives the degree of space non-locality in the
interaction, and which can be measured by locating the crossover to power law
scaling with respect to $|y-y'|$, in $\langle |\Psi_{ky}|^2|\Psi_{k'y'}|^2\rangle$.
\vskip 5pt
If $\bar z^3/\Delta z^2$ is large enough, there is a range of scale separations:
$\ln k'/k<\bar z^3/\Delta z^2$ where discreteness effects are relevant. 
In this regime, only one term contributes in the sums over $m$ and $n$ in Eqn. (9): the one
corresponding to the closest $\bar k_n$ and $\bar k_m$ respectively to $\bar k$ and $\bar k'$.
(The sum over $p$ remains continuous, provided $kL$ is large). Setting again $a_A=a_S$, the 
result, after some lengthy algebra, is equivalent to the one of Eqn. (11):
$$
\langle |\Psi_{ky}|^2|\Psi_{k'y'}|^2\rangle_c\simeq
{2c_4\pi^{3\over 2}\bar z^{1\over 2}\over\Delta z\hat ba_S}
\int_0^{\ln kL}\d x\,H^{-{1\over 2}}
\Big(1+F\e^{-2x}\Big)^{-{1\over 2}}
$$
$$
\times\exp\Big[ (\zeta'_4-2\zeta'_2)x-
\Big(\e^{2x}+F\Big)^{-1}{k^2|y-y'|^2\over\hat b^2a_S^2}
-{\bar z^3\Delta_{kk'}^2\over 2H\Delta z^2}\Big]
\eqno(13)
$$
where $H=2x+\ln k'/k+{4\bar z^2\over\Delta z^2a_S^2}$ and $\Delta_{kk'}=\bar z^{-1}\ln k'/k-
{\rm int}(\bar z^{-1}\ln k'/k)$ is the decimal part of $\bar z^{-1}\ln k'/k$. The 
main difference from Eqn. (11) lies in the quadratic term in $\Delta_{kk'}$ in the
exponential. As in the case of Eqn. (11), the integral receives contribution for
$\ln G<x<\ln kL$. The integrand has a saddle at:
$ \bar x\sim a_S(\bar z/2)^{3\over 2}(2\zeta_2-\zeta_4)^{-{1\over 2}} |\Delta_{kk'}|
-2\bar z^2 $ which starts playing a role, however, only at very small values of 
$\Delta_{kk'}$.  Thus the integral can be estimated almost always with steepest 
descent at $x\sim \ln G$. The result is:
$$
\langle |\Psi_{ky}|^2|\Psi_{k'y'}|^2\rangle_c\simeq
{2c_4\pi^{3\over 2}\bar z^{1\over 2}\over\Delta z\hat ba_S}(kL)^{-\zeta'_4}(k'/k)^{-\zeta'_2}
G^{2\zeta'_2-\zeta'_4}
$$
$$
\times
\max\Big(\ln GkL,(2\zeta'_2-\zeta'_4)^{-1}\Big)
\exp\Big(-{\bar z^3\Delta_{kk'}^2\over 2\Delta z^2
(\ln G+({2\bar z\over a_S\Delta z})^2)}\Big)
\eqno(14)
$$
which differs from Eqn. (12), again  because of the term in $\Delta_{kk'}$. This produces oscillations
in the correlation dependence on $\ln k'/k$, (lacunarity),
which have period $\bar z$, and die when
$\ln k'/k$ or $\ln{k|y-y'|\over\hat ba_S}$ become larger than 
$\bar z^3\Delta z^{-2}$. This at least in the physically interesting regime where the uncertainty
in the wavevector over a cascade step is of the same order or larger than the spectral
width of $w_S$. 

Dynamically, what happens is that the correlation between $w_S$ at a scale
separation $\ln k'/k$ which is not an integer number of $\bar z$, is due to a 
common ancestor at a scale $\hat k$, which must be distant a sufficient number $n$ of 
cascade steps from $k$. This is necessary for the width 
$n^{1\over 2}\Delta z$ to cover the difference $\Delta_{kk'}$ between 
$\bar z^{-1}\ln k'/k$ and the next integer. Clearly, the greater the number of cascade
steps, the smaller the contribution to the correlation, and this explains the decrease in
$\Delta_{kk'}$, of $\langle |\Psi_{ky}|^2|\Psi_{k'y'}|^2\rangle_c$, shown in Eqns. 
(13-14). 

We thus have three additional characteristic parameters describing the structure 
of a cascade generated random field: the oscillation period in $\ln k'/k$ of 
the correlation, and the separation in space and scale over which these 
oscillations die off. As regards turbulence modelling, it is worth mentioning, that  
different values of $\bar z$ correspond to different degrees 
of non-locality in  the Fourier structure of the Navier Stokes equation energy 
transfer terms. Analysis carried on by means of direct numerical 
simulations and closure has shown indeed that such non-local effects may be an
important component of turbulence dynamics \cite{domaradzkii90}.

Summarizing, for any fluctuation scale $k$ we have up to five additional scales at our disposal, 
to characterize an intermittent random field. This beyond the standard spectrum
of scaling exponents, provided by structure function analysis.
It is worth considering that these quantities have been introduced through an operative definition, 
which allows their measurement independently of the signal being synthetic, and originating from a
random cascade. In particular, even in the case of a generic random field, in which one would 
expect 
the coexistence of fluctuations of different shapes, a concept like the coherence lenght 
$\lambda_S$ is going to maintain its physical meaning, at least on an average sense. The same 
holds for all the quantities obtained from the multiscale-multispace
correlation $\langle |\Psi_{ky}|^2|\Psi_{k'y'}|^2\rangle$:
the degree of space non-locality, 
identified by the crossover to scaling in the space separation, the lacunarity period, measured
through the oscillation in $\ln k'/k$, and the separation in scale and space for the decay of 
these oscillations. This suggests, that methods for the characterization of intermittent signals, 
like the ones presented here, could be of some interest also for people with access to experimental
high Reynolds numbers turbulence data.
\vskip 5pt
From the point of view of turbulence modelling, it is interesting to examine
in more detail what happens in some limiting cases.

A first limit is obtained when $\hat b=L/\lambda_S$, and corresponds to the
maximum degree of space non-locality in the interaction. In this regime, $F\sim  
(\lambda_S/L)^2$ and $\langle |\Psi_{ky}|^2|\Psi_{k'y'}|^2\rangle_c\sim (kk')^{-\zeta'_2}$,
which means that the wavelets $w_S$ are essentially uncorrelated, although intermittently
distributed; in fact, the one-eddy contribution leads still to the anomalous structure function 
scaling: $\langle |\Psi_{ky}|^4\rangle\sim (kL)^{-\zeta'_4}$. This situation corresponds to the
random eddy model, studied in \cite{olla97}.

A different, more interesting kind of space non-locality is obtained when $\lambda_S\sim L$,
which implies that $a_S=k\lambda_S\to\infty$ as $k\to\infty$. In this case, the building blocks 
of the random field are not wavelets anymore, but Fourier components.
We would have then a signal generated by a superposition of random phase
Fourier modes, analogous to the turbulence  picture one gets out of
statistical closures. The only difference would be the increasing intermittency of the
Fourier amplitudes as $k\to\infty$. However, as we could expect, this does not
produce any intermittency in the signal; in fact, it can be shown, by generalizing the result
of Eqns. (8), (12) and (14), to the case of $a_A$ fixed and $a_S$ large, that: 
$\langle |\Psi_{ky}|^2\rangle^{-2}\langle |\Psi_{ky}|^4\rangle_c\propto a_S^{-1}$. Hence, the
non-Gaussian contribution to the correlation decays like $k^{-1}$ as $k\to\infty$.

\vskip 10pt
\noindent{\bf Aknowledgements}: One of us (P.O.) would like to thank Jean-Fran\c cois Pinton, Sergio Ciliberto and Jens Eggers for interesting and stimulating discussion.

\end{document}